# Formation and Migration of Trans-Neptunian Objects

S.I. Ipatov

*George Mason University, VA, USA; Institute of Applied Mathematics, Moscow, Russia*

**Abstract.** Trans-Neptunian objects (TNOs) with diameter $d>100$ km moving now in not very eccentric orbits could be formed directly by the compression of large rarefied dust condensations (with semi-major axes $a>30$ AU), but not by the accretion of smaller solid planetesimals. A considerable portion of TNO binaries could be formed at the stage of compression of condensations. Five years before the first TNO was observed, we supposed that, besides TNOs formed beyond 30 AU and moving in low eccentric orbits, there were former planetesimals from the zone of the giant planets in highly eccentric orbits beyond Neptune. For the present mass of the trans-Neptunian belt, the collisional lifetime of 1-km TNO is about the age of the solar system. At the present time TNOs can supply a large amount of matter to the near-Earth space.

## FORMATION OF TRANS-NEPTUNIAN OBJECTS

The total mass of the present Edgeworth–Kuiper belt (EKB) for objects with $30 \leq a \leq 50$ AU is estimated [1, 2] to be about $(0.06–0.25)m_\oplus$, where $m_\oplus$ is the mass of the Earth. Objects moving in highly eccentric orbits (mainly with $a>50$ AU) are called "scattered disk objects" (SDOs). The total mass of SDOs in eccentric orbits between 40 and 200 AU has been estimated to be about $0.05m_\oplus$ [3] or $0.5m_\oplus$ [4].

It was considered by many authors that a dust disk around the forming Sun became thinner until its density reached a critical value about equal to the Roche density. At this density, the disk became unstable to perturbations by its own self-gravity and developed dust condensations. These initial condensations coagulated under collisions and formed larger condensations, which compressed and formed solid planetesimals. In [5] it was considered that initial dimensions of planetesimals in the zone of Neptune were about 100 km, and in the terrestrial feeding zone they were about 1 km. According to [6], the mass of the largest condensation in the region of Neptune could exceed $2m_\oplus$. Formation and collisional evolution of the EKB were investigated in [7–10]. In these models, the process of accumulation of trans-Neptunian objects (TNOs) took place at small (∼0.001) eccentricities and a massive belt. More references on the above problems are presented in [11, 12].

Our runs showed [11, 13] that maximum eccentricities of TNOs always exceed 0.05 during 20 Myr under the gravitational influence of the giant planets. Gas drag could decrease eccentricities of planetesimals, and the gravitational influence of the forming giant planets could be less than that of the present planets. Nevertheless, in our opinion, it is probable that, due to the gravitational influence of the forming giant planets, migrating planetesimals, and other TNOs, small eccentricities of TNOs could not exist during all the time needed for the accumulation of TNOs with diameter $d>100$ km.

Eneev [14] supposed that large TNOs and all planets were formed by compression of large rarefied dust–gas condensations. We do not think that planets could be formed in such a way, but we consider [12] that TNOs with $d\geq100$ km moving now in not very eccentric orbits could be formed directly by the compression of large rarefied dust condensations (with $a>30$ AU), but not by the accretion of smaller solid planetesimals. The role of turbulence could decrease with an increase of distance from the Sun, so, probably, condensations could be formed at least beyond Saturn's orbit.

Probably, some planetesimals with $d\sim100$–$1000$ km in the feeding zone of the giant planets and even some large main-belt asteroids also could be formed directly by the compression of rarefied dust condensations. Some smaller objects (TNOs, planetesimals, asteroids) could be debris of larger objects, and other such objects could be formed directly by compression of condensations. Even if at some instant of time at approximately the same distance from the Sun, the masses of initial condensations, which had been formed from the dust layer due to gravitational instability, had been almost identical, there was a distribution in masses of final condensations, which compressed into the planetesimals. As in the case of accumulation of planetesimals, there could be a "run–away" accretion of condensations. It may be possible that, during the time needed for compression of condensations into planetesimals, some largest final condensations could reach such masses that they formed planetesimals with diameter equal to several hundreds kilometers.

It is considered that TNO binaries can be produced due to the gravitational interactions or collisions of future binaries with an object (or objects) that entered their Hill sphere. In our opinion, binary TNOs (including Pluto–Charon) were probably formed at that time when heliocentric orbits of TNOs were almost circular. For such orbits, two TNOs entering inside their Hill sphere could move there for a long time (e.g., greater than half an orbital period [15]). We suppose that a considerable portion of TNO binaries could be formed at the stage of compression of condensations. At this stage, the diameters of condensations, and so probabilities of their mutual collisions and probabilities of formation of binaries were much greater than those for solid TNOs. The stage of condensations was longer for TNOs than that for asteroids, and therefore binary asteroids (which could be mainly formed after formation of solid objects) are less frequent and more differ in mass than binary TNOs. Besides, at the initial stage of solar system formation, eccentricities of asteroids could be mainly greater (due to the influence of the forming Jupiter and planetesimals from its feeding zone) than those for TNOs.

Five years before the first TNO was discovered in 1992, based on our runs of the formation of the giant planets we supposed [15] that there were two groups of TNOs and, besides TNOs formed beyond 30 AU and moving in low eccentric orbits, there were former planetesimals from the zone of the giant planets in highly eccentric orbits beyond Neptune. During accumulation of the giant planets, planetesimals with a total mass equal to several tens $m_\oplus$ could enter from the feeding zone of the giant planets into the trans-Neptunian region, increased eccentricities and inclinations of 'local' TNOs, which initial mass could exceed $10m_\oplus$, and swept most of them [15, 16]. A very small fraction of such planetesimals could be left in eccentrical orbits beyond Neptune and became SDOs.

The total mass of planetesimals in the feeding zones of the giant planets, probably,

didn't exceed $300m_\oplus$, and only a smaller part of them could get into the Oort and Hills clouds and into the region between 50 and 1000 AU. So it seems more probable that the total mass of the objects located beyond Neptune's orbit doesn't exceed several tens $m_\oplus$.

Our computer runs [16–18], in which gravitational interactions of bodies were taken into account with the use of the spheres method, showed that the embryos of Uranus and Neptune could increase their semi-major axes from <10 AU to their present values, moving permanently in orbits with small eccentricities, due to gravitational interactions with the migrating planetesimals. Later on, Thommes et al. [19, 20] obtained similar results using direct numerical integration. The comparison of the results presented in [16–20] shows that the method of spheres can provide statistically reliable results for many bodies moving in eccentric orbits.

## COLLISIONAL EVOLUTION OF TRANS-NEPTUNIAN OBJECTS

Our estimates [11, 21] of the frequency of collisions of bodies in the present EKB and in the main asteroid belt (MAB) are of the same order of magnitude as the estimates obtained by other scientists (e.g., in [10]). For the EKB with a total mass $M_{EKB} \sim 0.1 m_\oplus$ and the ratio $s$ of masses of two colliding bodies, for which a collisional destruction of a larger body usually takes place, equal to $10^3$, ($s$ depends on composition and diameters of objects, a collisional specific energy, and collisional velocity) a collisional lifetime $T_c$ of a body with $d$=100 km is about 30 Gyr [13]. For $10^{12}$ 100-m TNOs, 1-km TNO collides with one of 100-m TNOs on average ones in 3 Gyr. At the same $s$, the values of $T_c$ for 1-km TNOs are of the same order of magnitude as those for main-belt asteroids.

The mean energy of a collision is proportional to $v_c^2$, where $v_c$ is the relative velocity of a collision. For the EKB the mean energy of a collision and, for the same composition of two colliding bodies, also the ratio $s$ needed for destruction of a larger colliding body in the EKB are smaller by about a factor of $k \approx 20$ than those for the MAB. However, as it can be more easy to destruct icy TNOs than rocky bodies in the MAB, then $s$ can be much larger for the EKB, and collisional lifetimes of small bodies in the EKB can be of the same order as those in the MAB. If some TNOs are porous, then it may be more difficult to destroy them than icy and even rocky bodies and their collisional lifetimes can be larger than those for main-belt asteroids of the same sizes.

The total mass of SDOs moving in highly eccentric orbits between 40 and 200 AU is considered to be of the same order or greater than $M_{EKB}$. The mean energy of a collision of a SDO with a TNO is greater than that for two colliding TNOs of the same masses. Therefore, though SDOs spend a smaller part of their lifetimes at a distance $R$<50 AU, the probability of a destruction of a TNO (with 30<$a$<50 AU) by SDOs can be of the same order of magnitude as that by TNOs. In [11, 21] we also investigated the orbital evolution of TNOs under the mutual gravitational influence.

The total mass of planetesimals that entered the trans-Neptunian region during the formation of the giant planets could be equal to several tens $m_\oplus$, and this time interval could be about several tens Myr. Besides, the initial mass of the EKB can be much larger ($\geq 10 m_\oplus$) than its present mass. Therefore, TNOs could be more often destroyed during planet formation than during last 4 Gyr.

# MIGRATION OF TRANS-NEPTUNIAN OBJECTS

As migration of TNOs to Jupiter's orbit was investigated by several authors, we [22–24] have made a series of simulations of the orbital evolution of 25,000 Jupiter-crossing objects (JCOs) under the gravitational influence of planets. Our runs showed that if one observes former comets in near-Earth object (NEO) orbits, then most of them could have already moved in such orbits for millions of years. Results of our runs testify in favor of at least one of these conclusions: 1) the portion of 1-km former TNOs among NEOs can exceed several tens of percents, 2) the number of TNOs migrating inside the solar system can be smaller by a factor of several than it was earlier considered, 3) most of 1-km former TNOs that had got NEO orbits disintegrated into mini-comets and dust during a smaller part of their dynamical lifetimes if these lifetimes are not small.

## ACKNOWLEDGMENTS

This work was supported by NASA (NAG5-10776), INTAS (00-240) and RFBR (01-02-17540).